\documentclass[10pt,aps,prl,twocolumn,amsmath,amssymb,superscriptaddress]{revtex4-1}

\usepackage{amsmath}
\usepackage{graphicx}
\usepackage{color}
\usepackage{multirow}
\usepackage[caption = false]{subfig}
\usepackage{array}
\newcolumntype{P}[1]{>{\centering\arraybackslash}p{#1}}
\newcolumntype{L}[1]{>{\raggedright\arraybackslash}p{#1}}
\newcolumntype{R}[1]{>{\raggedleft\arraybackslash}p{#1}}
\def\Nd{{N_d}}
\def\Nddiff{{N_{d}^{\mathrm{diff}}}}

\begin{document}
\renewcommand{\arraystretch}{1.2}

\title{Stochastic perturbation theory to correct non-linearly parametrized wavefunctions}
\author{Sandeep Sharma}
\email{sanshar@gmail.com}
\affiliation{Department of Chemistry and Biochemistry, University of Colorado Boulder, Boulder, CO 80302, USA}
\begin{abstract}
We introduce an algorithm that can be used to perform stochastic perturbation theory (sPT) to correct any non-linearly parametrized wavefunction that can be optimized using orbital space Variational Monte Carlo (VMC). Although the variational method gaurantees that the VMC energy can be systematically improved the cost of doing so in practice is often prohibitive. The sPT algorithm presented in this work represents an efficient way to improve the VMC energies with a relatively small computational overhead. We demonstrate that for the carbon dimer and Fe-porphyrin the sPT algorithm is able to capture $>97\%$ and $>60\%$  respectively of the correlation energy missing from the zeroth order wavefunction. Further, the sPT algorithm is also ideally suited for massively parallel computations because it delivers super-linear speedup with an increasing number of processors.
\end{abstract}
\maketitle

In the variational method, approximate ground or excited state energies and wavefunctions are obtained by varying the parameters of the wavefunction to minimize its energy. The energy so obtained converges to the exact value from above as the flexibility of the wavefunction is enhanced by increasing the number of adjustable parameters. The simplest such wavefunction is the linearly parametrized configuration interaction (CI) wavefunction $|\Psi\rangle = \sum_i c_i |D_i\rangle$, where $c_i$ are the adjustable parameters. In the limit that the summation is over all determinants of the Hilbert space we obtain the full configuration interaction (FCI) wavefunction that is exact for a given Hilbert space. However, FCI suffers from the fact that only systems containing about 20 electrons in 20 orbitals can be treated because the size of the Hilbert space increases exponentially with the number of electrons. This bottleneck can be overcome by using a non-linearly parametrized wavefunction which allows one to make judicious use of the physics of the problem, e.g. the matrix product state (MPS) wavefunction\cite{Ostlund1995,Schollwock2011}, which is variationally optimized using the density matrix renormalization group (DMRG) algorithm\cite{white1992density,white1993density}, is well suited to strongly correlated one-dimensional systems. Other examples include the Jastrow based states\cite{Eric2013}, Resonating valence bond\cite{Sorella07}, nonorthogonal configuration interaction\cite{Sundstrom14} and neural network quantum states\cite{Carleo2017} etc. 

Often, it is not possible to obtain a polynomial scaling analytic expression to calculate the energies of such wavefunctions ($E= \frac{\langle \Psi|\hat{H}|\Psi\rangle}{\langle\Psi|\Psi\rangle}$) and instead one has to resort to the Variational Monte-Carlo (VMC) method to obtain the energy and subsequently optimize it. Two major drawbacks of the VMC algorithms are, (i) attempts to make the wavefunction more accurate by increasing the number of parameters often leads to high redundancy as measured by the condition number of the Hessian matrix, $K_{ij} = \frac{\partial^2 E}{\partial p_i\partial p_j}$, where $p_i$ and $p_j$ are wavefunction parameters and (ii) the commonly used algorithm to minimize the energy called the linear method\cite{UmrTouFilSorHen-PRL-07,TouUmr-JCP-08} requires that the Hessian matrix be stored in memory, limiting the number of parameters that can be included in the wavefunction. Due to the difficulties posed by the last two points, the accuracy of the VMC wavefunctions is limited. In this work we restrict ourselves to the orbital VMC wavefunctions which are designed to deliver a variational energy in a basis set. Unlike their real space counterparts -- which are routinely corrected by performing a followup diffusion Monte-Carlo (DMC) calculation -- the orbital space DMC method, also referred to as Green's function Monte-Carlo (GFMC)\cite{Haaf1995,Sorella98}, is much less widely used due to its high cost. 
    
In this report, we will show that after an approximate wavefunction has been obtained using the variational method, it can be made more accurate by using perturbation theory. Unlike, the wavefunction itself, the perturbative correction to the wavefunction is written using a linear parametrization of the type used in CI wavefunction, which eliminates the difficulties associated with the non-linear parametrization. Further, the perturbation theory will be performed using a stochastic algorithm which overcomes the need to store all the coefficients in memory simultaneously. 
 Such a stochastic perturbation theory (sPT) algorithm was recently proposed to correct an approximate CI wavefunction obtained by the semistochastic heat-bath configuration interaction (SHCI) algorithm\cite{HolChaUmr-JCTC-16,ShaHolUmr-JCTC-17} (more details are given in the next paragraph). Here, we show that the sPT algorithm can be extended to correct any wavefunction including the types of non-linearly parametrized wavefunction used in VMC. The algorithm presented here has several attractive features: the memory cost of performing the sPT algorithm can be made arbitrarily small, the sPT algorithm provides a way to improve the energies of the VMC wavefunctions without having to incur the expense of increasing the number of parameters in the wavefunction and finally, super-linear speedup of computation with the number of processors is observed.
Although the algorithm can be used with virtually any VMC wavefunction, here we demonstrate it using MPS; extension of the algorithm to other wavefunction ansatz is underway. 

Before giving the details of the current algorithm the original sPT algorithm used to correct the CI wavefunction obtained from the SHCI algorithm is presented. (SHCI algorithm is an instance of a more general class of methods called selected configuration interaction followed by perturbation theory (SCI-PT)\cite{Huron1973,Buenker1974}) Only the salient features of the SHCI algorithm will be described here and the reader is referred to the original publication\cite{HolChaUmr-JCTC-16,ShaHolUmr-JCTC-17} for further details. SHCI consists of two steps, in the first step a set of ``important" determinants ($\mathcal{V}$) are selected and the wavefunction 
$|\Psi\rangle = \sum_{i \in \mathcal{V}} c_i |D_i\rangle$
is obtained by diagonalizing the Hamiltonian in the subspace of these determinants. In the second step, the correction to the energy and wavefunction due to all determinants is calculated by performing a perturbation theory in which the zeroth order Hamiltonian is
\begin{align}
\hat{H}_0 = \sum_{i,j \in \mathcal{V}} H_{ij} |D_i\rangle\langle D_j| + \sum_{a \notin \mathcal{V}} H_{aa} |D_a\rangle\langle D_a|, \label{eq:sciH0}
\end{align}
where $H_{ij} = \langle D_i|\hat{H}|D_j\rangle$ is the Hamiltonian matrix elements. The choice of $\hat{H}_0$ is prompted by the fact that it is diagonal in the space of determinants not present in $\mathcal{V}$, which ensures that the first order correction to the wavefunction 
\begin{align}
 |\Psi_1\rangle &= \hat{Q}\frac{1}{E_0- H_0}\hat{Q}V|\Psi_0\rangle \label{eq:psi1}\\
 & = \sum_{a\notin \mathcal{V}} \frac{\sum_{i\in \mathcal{V}}H_{ai} c_i}{E_0 - E_a}  |D_a\rangle,
 \end{align}
 and the second order correction to the energy 
 \begin{align}
 \Delta E_{2} & = \langle\Psi_0|V|\Psi_1\rangle =\sum_{a\notin \mathcal{V}} \frac{\left(\sum_{i\in \mathcal{V}}H_{ai} c_i\right)^2}{E_0 - E_a}\label{eq:PTb} ,
 \end{align}
can be calculated without having to invert a matrix. Here, $\hat{Q} = 1 - |\Psi_0\rangle\langle\Psi_0|$ guarantees that the first order wavefunction is orthogonal to the zeroth order wavefunction. The orthogonality is trivially satisfied because only determinants not present in $\mathcal{V}$ contribute to the first order correction.

The second order energy can be calculated efficiently by first looping over all the determinants $|D_i\rangle$ in $\mathcal{V}$ and calculating $H_{ai}c_i$ for all the determinants $|D_a\rangle$ ($\notin \mathcal{V}$) which have a non-zero Hamiltonian transition matrix element $H_{ai} \neq 0$. Then the determinants and the corresponding elements $H_{ai}c_i$ are sorted by the determinants $|D_a\rangle$ and finally all the contributions $H_{ai}c_i$ for $|D_a\rangle$ are accumulated and squared to evaluate the numerator of Equation~\ref{eq:PTb}. This procedure allows one to calculate the second order correction at a cost that scales as $O(N\log N)$ ($O(N)$ by using a Hash function instead of sorting), where $N$ is equal to the number of determinants $|D_a\rangle$. 

The drawback of the algorithm is that all the $N$ determinants need to be stored in memory simultaneously which restricts the size of the problem that can be solved. The sPT algorithm introduced in\cite{ShaHolUmr-JCTC-17}, overcomes this requirement by evaluation Equation~\ref{eq:PTb} using a Monte-Carlo algorithm, with the result that the memory bottleneck can be completely eliminated at the cost of introducing a stochastic error. Each Monte-Carlo iteration uses essentially the same algorithm described above with the important difference that one can use as few as two determinants sampled from the variational space, making each iteration nearly as cheap as a second order M\"oller-Plessett calculation.  

The second order correction to the energy 
\begin{align}
 \Delta E_{2} & = \langle\Psi_0|V|\Psi_1\rangle= \sum_{a} \frac{\sum_{ij}H_{ai}H_{aj} c_ic_j}{E_0 - E_a}\label{eq:1}
\end{align}
is a bilinear function in the coefficients in the zeroth order state. In each Monte-Carlo iteration a set of $N_d$ determinants from the space $\mathcal{V}$ are chosen such that each determinant $|D_i\rangle$ has a non-zero probability $p_i$ of being selected. This will result in a selection of $N_d^{\mathrm{diff}}$ unique determinants with each determinant appearing $w_i$ times, such that
$\sum_i^{N_d^{\mathrm{diff}}} w_i = N_d$.
By using the fact that the numbers $w_i$ are distributed according to the well known multinomial distribution, one can show that the unbiased estimate of the second order energy can be evaluated as
\begin{align}
\Delta E_{2}=& \frac{1}{\Nd(\Nd-1)} \left\langle \sum_{a} \frac{1}{E_0 - E_a} \left[\left(\sum_{i}^{\Nddiff} \frac{ w_i  c_i H_{ai}}{p_i}\right)^2  \right.\right.\nonumber\\
&\left.\left.+\sum_{i}^{\Nddiff} \left(\frac{w_i(\Nd-1)}{p_i } - \frac{ w_i^2}{p_i^2}\right)c_i^2 H_{ai}^2\right] \right\rangle, \label{eq:stoc}
\end{align}
where all summations in the third line are over just $\Nddiff$ determinants. The rate of convergence of the stochastic evaluation of the energy will depend on the values $p_i$. Although, the optimal choice of $p_i$ is not known, we find that using $p_i = \frac{|c_i|}{\sum_i |c_i|}$ ($c_i$ is the amplitude of the determinant $D_i$) delivers rapid convergence. We next describe how this algorithm can be extended to situations where the zeroth order wavefunction is non-linearly parametrized.

We start by assuming that the DMRG algorithm is used to obtain the zeroth order energy and zeroth order wavefunction which is an MPS of bond dimension $M$. The second order correction $E_2$ to the wavefunction $|\Psi_0\rangle$, with energy $E_0$ is given by,
\begin{align}
E_2 = \langle \Psi_0|\hat{V}\hat{Q}\frac{1}{\hat{H}_0 - E_0} \hat{Q}\hat{V}|\Psi_0\rangle,\label{eq:e2}
\end{align}
where, as before, $\hat{H}_0$ and $\hat{V}$ are respectively the zeroth order Hamiltonian and perturbation respectively, such that $\hat{H} = \hat{H}_0 + \hat{V}$, and $\hat{Q} = 1 - |\Psi_0\rangle\langle\Psi_0|$ is the projector onto the space complementary to $|\Psi_0\rangle$.  A non-linearly parametrized wavefunction such as an MPS introduces difficulties that were not present in the linearly parametrized zeroth order wavefunction. First, because the zeroth order wavefunction will potentially have non-zero overlap with all determinants in the Hilbert space, no nontrivial $\hat{H}_0$ can be defined that simultaneously has the zeroth order wavefunction and also all the determinants that contribute to the first order correction as its eigenfunctions. Second, the action of the projector $\hat{Q}$ to ensure that the first order wavefunction is orthogonal to the zeroth order wavefunction is no longer trivial.

The Hamiltonian is partitioned by defining 
\begin{align}
\hat{H}_0 = \hat{P}\hat{H}\hat{P} + \hat{Q}\hat{H}^{\mathrm{EN}} \hat{Q}, \label{eq:h0}
\end{align}
where, 
\begin{align}
\hat{H}^{\mathrm{EN}} =& \sum_i t_{ii} a_i^\dag a_i + \sum_{ij} \langle ij || ij\rangle a_i^\dag a_j^\dag a_j a_i, 
\end{align}
is the Esptein-Nesbet Hamiltonian\cite{Kvasnicka1976} and $ \langle ij || ij\rangle =  \langle ij | ij\rangle - \langle ij | ji\rangle$ is the antisymmetrized two electron integral. This partitioning ensure that the zeroth order wavefunction and all the determinants that have a zero overlap with $\Psi_0$ are eigenfunctions of $\hat{H}_0$. Use of this Hamiltonian instead of the one defined in Equation~\ref{eq:sciH0} for the special case of the selected CI wavefunction will give the same second order energy correction. In order to perform stochastic perturbation theory with an MPS and $\hat{H}_0$ defined in Equation~\ref{eq:h0} we will describe the three steps of the algorithm that make it different from the SHCI algorithm:
\begin{itemize}
\item application of $\hat{Q}\hat{V}$ on the zeroth order wavefunction
\item action of $\frac{1}{\hat{H}_0 - E_0}$ without having to invert a large matrix and
\item sampling determinants $|D_i\rangle$ that have a non-zero overlap with $|\Psi_0\rangle$ for use in Equation~\ref{eq:stoc},
\end{itemize}

The first step of the algorithm involves acting the operator $\hat{Q}\hat{V}$ on the sampled wavefunction $|\Psi_0\rangle$. Application of the projection operator $\hat{Q}$ to ensure the orthogonality of the first order wavefunction to $|\Psi_0\rangle$ (see Equation~\ref{eq:psi1}) is not straight forward, because unlike in SHCI, the determinants cannot be clearly partitioned into $\mathcal{V}$ that contribute to $|\Psi_0\rangle$ and all other determinants that contribute to $|\Psi_1\rangle$. To overcome this difficulty we use the following identity 
\begin{align}
\hat{Q}\hat{V}|\Psi_0\rangle =& (\hat{H} - \hat{H_0})|\Psi_0\rangle - \hat{P}(\hat{H} - \hat{H_0})|\Psi_0\rangle \\
&= (\hat{H} - E_0 )|\Psi_0\rangle.\label{eq:trick1}
\end{align}
Thus given the $\hat{H}_0$ in Equation~\ref{eq:h0}, the action of $(\hat{H} - E_0)$ is equivalent to that of $\hat{Q}\hat{V}$ on $|\Psi_0\rangle$. 
 
 Second, unlike in the case of SHCI, the $\hat{H}_0$ defined in Equation~\ref{eq:h0} is not diagonal in the determinantal basis, thus inverting $\hat{H}_0 - E_0$ is no longer a trivial operation and can be as computationally expensive as diagonalizing $\hat{H}$. To overcome this difficulty we use the approximation
 \begin{align}
\langle D_i|\frac{1}{\hat{H}_0 - E_0}|D_j \rangle \approx \langle D_i|\frac{1}{\hat{H}^{\mathrm{EN}} - E_0}|D_j \rangle= \frac{\delta_{ij}}{E_i - E_0},\label{eq:approx}
 \end{align}
where $E_i$ is the energy of the determinant $|D_i\rangle$. This might appear as a large uncontrolled approximation but in practice it leads to small errors as we will demonstrate while discussing the results for the carbon dimer. It is worth remembering that this approximation is also used with much success in the Jacobi-Davidson algorithm to solve the eigenvalue problem.

With the help of Equation~\ref{eq:trick1} and the approximation in Equation~\ref{eq:approx}, the second order correction to the energy in Equation~\ref{eq:e2} can be written as
\begin{align}
E_2 = \langle \Psi_0|(\hat{H} - E_0 )\frac{1}{\hat{H}^{\mathrm{EN}} - E_0} (\hat{H} - E_0 )|\Psi_0\rangle.\label{eq:e2_2}
\end{align}

Finally, to evaluate the expression in Equation~\ref{eq:e2_2} one can use the replica method\cite{Overy14} according to which, on the $I^{th}$ iteration a bra state and a ket state are sampled independently from the zeroth order wavefunction to obtain $\langle \Psi_B^I| = \sum_i c_i^I \langle D_i|$ and $| \Psi_K^I\rangle = \sum_j d_j^I |D_j\rangle$, where the summation over $i$ and $j$ can be made as small as a single determinant. (To sample from the zeroth order state we regard it as a normalized probability distribution with each determinant $|D_i\rangle$ appearing with probability $p_i = \frac{|c_i|}{\sum_i |c_i|}$.) These sampled bra and ket states can then be used to get an unbiased estimate of the second order energy
\begin{align}
E_2^I =& \langle \Psi_B^I|(\hat{H} - E_0 )\frac{1}{\hat{H}^{\mathrm{EN}} - E_0} (\hat{H} - E_0 )|\Psi_K^J\rangle\\
=& \sum_{ij} c_i^Id_j^I  \langle D_i|(\hat{H} - E_0 )\frac{1}{\hat{H}^{\mathrm{EN}} - E_0} (\hat{H} - E_0 )|D_j\rangle \label{eq:5}
\end{align}
By averaging over a large number of iterations the stochastic error can be reduced to an acceptable value. The sampled wavefunctions $\langle \Psi_B^I|$ and $| \Psi_K^I\rangle$ can be straightforwardly obtained using the Metropolis algorithm as long as the overlap of a determinant with the zeroth order wavefunction can be calculated efficiently.

In this work, instead of using the replica method, we rely on the special property of the MPS that allows one to sample determinants directly without recourse to the Metropolis algorithm, thus eliminating autocorrelations in the Monte-Carlo simulation. This allows us to use Equation~\ref{eq:stoc} instead of Equation~\ref{eq:5} to evaluate the second order perturbation theory. The algorithm used for sampling is reminiscent of the one used to generate minimally entangled typical quantum states (METTS)\cite{White09} from an MPS with the key difference that while in METTS we sample determinants with a probability proportional to $|\langle D_i|\Psi\rangle|^2$, here we are trying to sample states with a probability proportional to $|\langle D_i|\Psi\rangle|$. Although an algorithm to sample determinants with this exact probability is unknown, below we outline an algorithm that is a reasonable approximation. To do this, one uses the sweep algorithm and at each sweep iteration, only one out of all the four possible site bases is retained during decimation. During the right sweep at site $n$, the current state can be written as
\begin{align}
|\Psi\rangle = \sum_{s_n^ir} C_{s_n^ir}|s_1\rangle |s_2\rangle\cdots |s_{n-1}\rangle  |s_n^i\rangle |r\rangle
\end{align} 
where $|s_1\rangle \cdots |s_{n-1}\rangle$ are the site states retained during the sweep for each of the first $n-1$ sites and summation is over the four possible site bases ($|s_n^i\rangle$) of the $n^{th}$ site and all possible right canonical states ($|r\rangle$). At this iteration the state $|s_n^i\rangle$ is retained with probability
\begin{align}
p^i_n = \frac{\sum_{r} |C_{s_n^ir}|}{\sum_{s_n^j r} |C_{s_n^j r}|},
\end{align}
and all the other states on site $n$ are discarded. At the end of the sweep a single determinant is obtained with a probability equal to $\prod_{\mathrm{sites~} n} |p^i_n|$. At each Monte-Carlo iteration, this calculation is performed $\Nd$ times to obtain $\Nd$ independently sampled determinants with possible repeats $w_i$. Once these determinants have been generated along with the probabilities, the expression in Equation~\ref{eq:stoc} can be directly used to obtain an unbiased estimate of the second order energy.
\begin{table}
\centering
\caption{DMRG energies of C$_2$ calculated with a series of increasing bond dimensions $M$. The ``exact" energy is calculated with $M=1500$. The sPT and MPSPT corrected energies are tabulated in the third and fourth column. The good agreement between the sPT and MPSPT energies indicates that the error introduced due to Equation~\ref{eq:approx} is small. Finally, the last column indicates the \% of the correlation energy recovered by the sPT algorithm that was missing from the zeroth order state.}\label{tab:c2}
\begin{tabular}{cccccccc}
\hline
\hline
$M$&	DMRG&~~&	sPT&~~&	MPSPT&~~& \% Corr. E.\\
       &                  &     &           &     &             &     &(sPT)\\
\hline
50&	-75.6829&	&-75.7307(1)&&	-75.7290&&98\\
100&	-75.7092&	&-75.7312(1)&&	-75.7299&&97\\
200&	-75.7219&	&-75.7317(1)&&	-75.7307&&98\\
400&	-75.7278&	&-75.7318(1)&&	-75.7311&&98\\
\hline
Exact~~&	-75.7319	&\\	
\hline
\end{tabular}
\end{table}
\begin{figure}
\begin{center}
\includegraphics[width=0.35\textwidth]{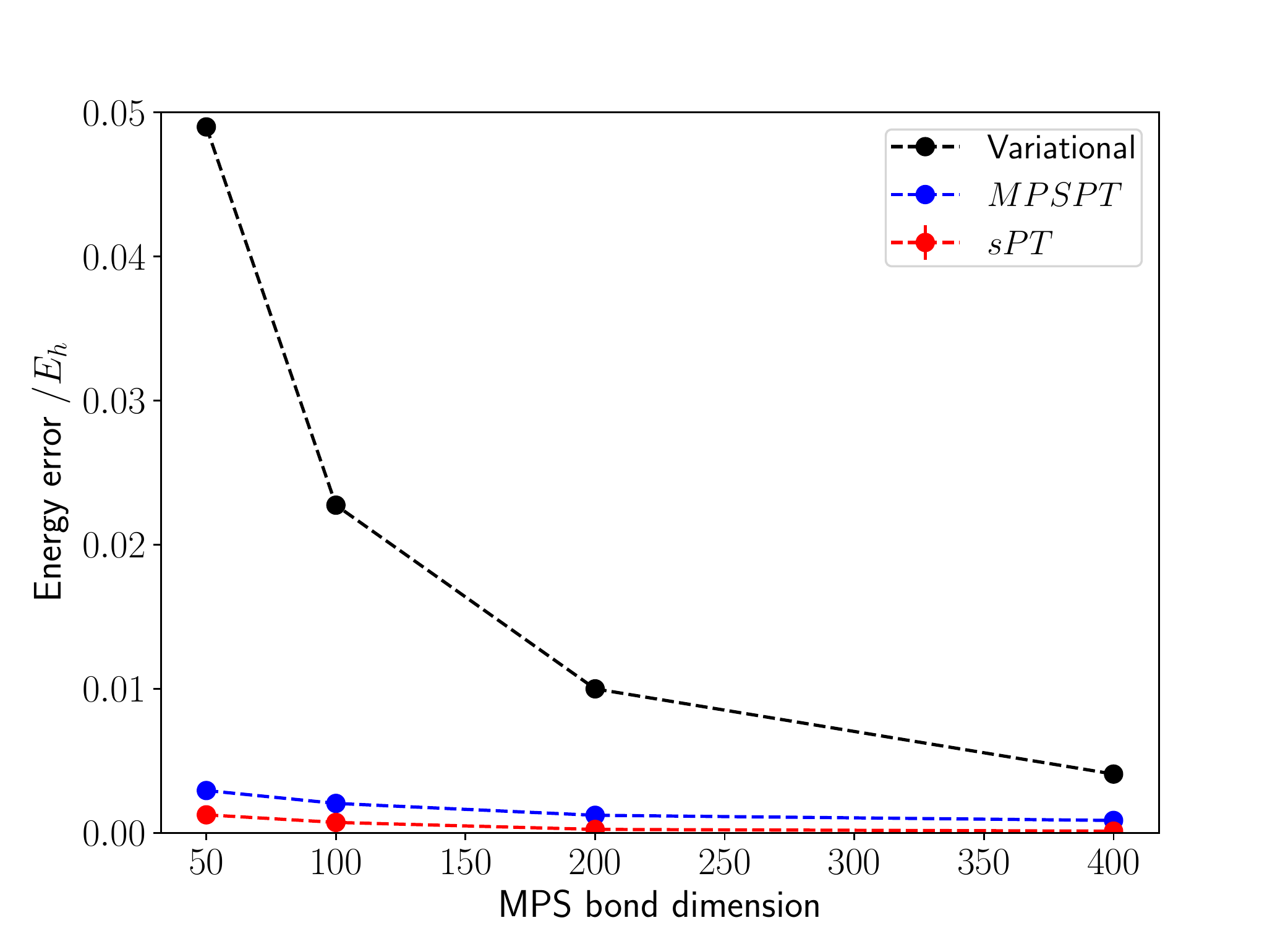}
\caption{The black line shows the error in DMRG energy of the ground state of the C$_2$ molecule as the MPS bond dimension is progressively increased. The blue and the red lines are respectively the errors after the second order sPT and MPSPT energies are used to correct the DMRG energy.}
\end{center}
\end{figure}

We test sPT algorithm on the Carbon dimer at its equilibrium bond length of 1.2425 \AA~with the cc-pVDZ basis set. An initial Hartree-Fock calculation was performed and subsequently, all 12 electrons were fully correlated in 28 canonical orbitals. The aim of these calculations is to test the accuracy of the approximation in Equation~\ref{eq:approx} and also to demonstrate the effectiveness of the sPT algorithm. An initial DMRG calculation was performed with a ``small" MPS bond dimension $M$ and subsequently the sPT algorithm outlined above was used to calculate the second order correction to the DMRG energy. In addition to the sPT algorithm, the exact second order energy was calculated using the matrix product state perturbation theory (MPSPT)\cite{Sharma2014a}. The MPSPT algorithm allows one to deterministically solve the linear equation in Equation~\ref{eq:psi1} for an arbitrary zeroth order Hamiltonian. In this algorithm, the first order correction $|\Psi_1\rangle$ is represented as an MPS and then sweep algorithm similar to that in DMRG are performed to solve the linear equation. By increasing the bond dimension of the MPS used to represent $|\Psi_1\rangle$ the exact second order energy can be calculated deterministically. We refer the reader to the original publication for further details, but for our purposes, it is sufficient to know that the second order correction can be calculated exactly, which when compared to the result of the sPT calculation allows us to gauge the accuracy of the approximation in Equation~\ref{eq:approx}. The results in Table~\ref{tab:c2} show that the sPT perturbation energies differ from the MPSPT energies by less than 2 mHa. This error decreases as the zeroth order wavefunction become more accurate and in the limit that it becomes the exact eigenstate of the Hamiltonian, the second order correction vanishes.
\begin{figure}
\begin{center}
\includegraphics[width=0.35\textwidth]{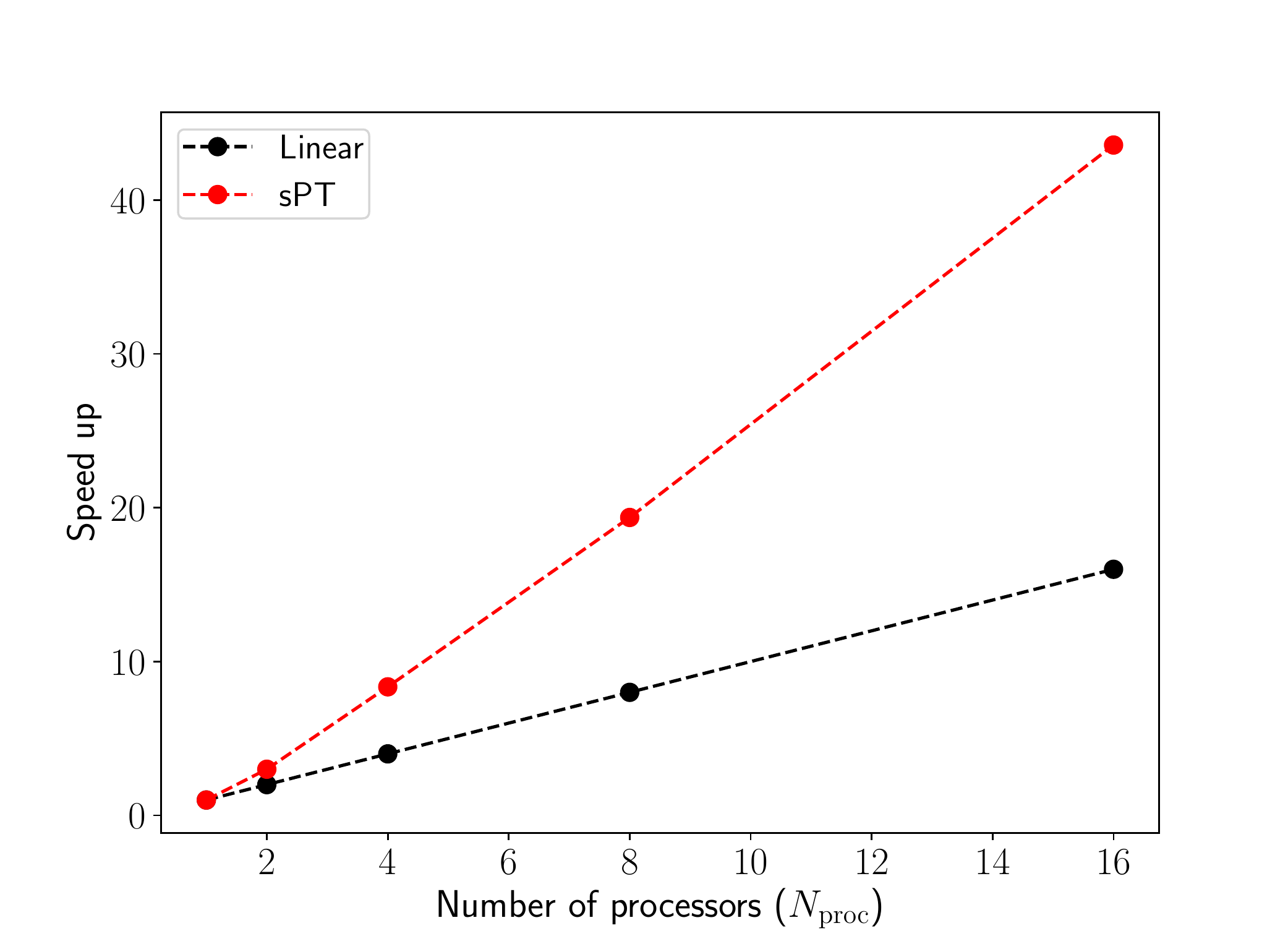}
\caption{The figure shows the speed up in the sPT computation as a function of the number of processors (in red). It also shows the ideal linear speed up (in black). All calculations were performed on the Carbon dimer and used an MPS of bond dimension $M=200$ as the zeroth order wavefunction. In each sPT iteration 100 determinants per processor were used to sample the MPS.}\label{fig:speedup}
\end{center}
\end{figure}
\begin{table}
\centering
\caption{DMRG energies of Fe-porphyrin calculated with a series of increasing bond dimensions $M$. The ``exact" energy is obtained from Ref.~\citenum{James17}. The sPT corrected energy is tabulated in the third column. The last column indicates the \% of the correlation energy recovered by the sPT algorithm that was missing from the zeroth order state.}\label{tab:fep}
\begin{tabular}{cccccc}
\hline
\hline
$M$&	DMRG&~~&	sPT&~~& \% Corr. E.\\
       &                  &     &           &    &(sPT)\\
\hline
100&	-2244.9901&&	-2245.0169(3)&&65	\\
200&	-2245.0041&&	-2245.0222(3)&&66	\\
400&	-2245.0131&&	-2245.0256(4)&&68	\\
800&	-2245.0203&&	-2245.0269(4)&&59	\\
1600&	-2245.0253&&	-2245.0291(5)&&62\\	
3200	&-2245.0283&&	-2245.0304(3)&&68	\\
\hline
Exact~~&	-2245.0314(5)&	&\\	
\hline
\end{tabular}
\end{table}

Another striking feature of the sPT algorithm is that better than linear speed up with the number of processors is obtained, as shown in Figure~\ref{fig:speedup}. This result is attained by assuming that a fixed amount of memory is available per processor which allows one to sample the wavefunction using a maximum number ($d_p$) of determinants per processor. An embarrassingly parallel algorithm is obtained if each processor is used to perform a separate sPT calculation and the results from all the processors are averaged. However, if determinants from each processor are accumulated prior to the sPT calculation, then one obtains better than a linear speed up. The reason for this can be understood by noting that the second order energy is a non-linear function of the zeroth order wavefunction and thus an increase in $\Nd = d_pN_p$ ($N_p$ is the number of processors) used to approximate the zeroth order wavefunction results in better than linear speed up.  

Next, we perform the sPT calculation for the more challenging Fe-porphyrin (Fe(P)) system, which is an active site in several biological metalloenzymes such as hemoglobin, myoglobin etc. In this calculation, 32 electrons were correlated in 29 orbitals (20 C 2$_{pz}$ orbitals, 4 N 2$_{pz}$ orbitals and 5 Fe 3$_{d}$ orbitals) obtained from an HCISCF calculation\cite{James17} with the cc-pVDZ basis set. The results in Table~\ref{tab:fep} demonstrate that the sPT calculations are able to recover a significant portion of correlation energy missing from the zeroth order energy. For example, an sPT calculation with an MPS of bond dimension of 50 is able to account for a greater correlation energy than an MPS of bond dimension 400. Given that the memory cost of the calculation increases as the second power of $M$, this represents greater than a factor of 64 reduction in memory usage. 

In this work, we present an algorithm to perform stochastic perturbation theory that can be used to correct any wavefunction that is amenable to being optimized using the variational Monte Carlo (VMC) algorithm. The algorithm has a low memory overhead and can be performed efficiently on massively parallel machines because it delivers super-linear speedup of computation with an increase in the number of processors. In the systems we have attempted here, the sPT algorithm can capture between 60\% to 97\% of the correlation energy missing from the zeroth order wavefunction. Although we have demonstrated the use of the algorithm with MPS, we expect it to be much more useful for other wavefunctions for which robust algorithms such as DMRG are not available to optimize a large number of wavefunction parameters.  

The author acknowledges useful discussions with Cyrus Umrigar and the financial support from the University of Colorado, Boulder through the startup package.

\end{document}